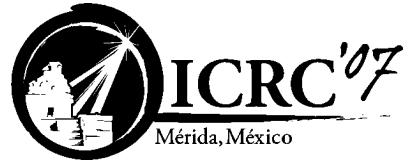

# The TrICE Prototype MAMPT Imaging Camera

K. BYRUM[1], J. CUNNINGHAM[4], G. DRAKE[1], E. HAYS[1,2], D. KIEDA[3], E. KOVACS[1], S. MAGILL[1,], L. NODULMAN[1], R. NORTHROP[2], S. SWORDY[2], R. WAGNER[1,], S.P. WAKELY[2], S.A. WISSEL[2]
[1]*Argonne National Laboratory*
[2]*University of Chicago, Enrico Fermi Institute*
[3]*University of Utah*
[4]*Loyola University*
contact byruml@anl.gov

**Abstract:** The Track Imaging Cerenkov Experiment (TrICE) is an air Cerenkov prototype telescope designed to use multi-anode photomultipliers to achieve a high angular resolution for measuring cosmic-ray composition at TeV-PeV energies. The TrICE camera, composed of 16 Hamamatsu R8900 16-channel multi-anode photomultiplier tubes, achieves 0.086 degree angular width per pixel over 1.5 degree wide field of view. We present a description of the TrICE camera design, calibration and performance.

## Introduction

The TrICE prototype [2] is designed to use a high resolution method for measuring TeV-PeV cosmic-ray composition by observation of the direct Cerenkov radiation from cosmic-ray nuclei before they interact in the atmosphere [1]. Angular resolution of the shower image is important for distinguishing the unique characteristics of the direct Cerenkov signal. Current Cerenkov imaging telescopes use conventional single-anode phototubes in the camera. TrICE improves the image angular resolution over existing telescopes by using multi-anode phototubes. An angular pixel size of .086 degrees in width is achieved by using Hamamatsu 16-channel R8900 multi-anode photomultipliers (MAPMTs). Here we describe the selection, testing, and performance of phototubes for TrICE, including modifications made to handle large currents from night sky background (NSB) light.

## The TrICE Camera Design

The TrICE camera consists of an array of 16 MAPMTs, 256 pixels total. The tubes are mounted on a circuit board that in turn mounts onto the imaging plane of the telescope. The camera incorporates a baffle to provide a shield from horizon light. Figure 1 shows the completed camera board populated with MAPMTs. Additional functionality on the circuit board allows for separate monitoring of the MAPMT currents. The analog signals are sent on short cables without pre-amplification to the electronics modules, where the integration of the current pulses and digitization of data are performed.

## MAPMT Requirements

The first task for designing the TrICE camera was selecting the appropriate photosensor. The optics allowed for an optimum sensor pitch of approximately 5 mm. The camera specifications on the photosensor were that the sensor could distinguish single photo-electrons (PEs), that it have low crosstalk and the ability to handle pulsed operation in the presence of the large DC anode currents from the NSB, that the sensor have good linear response over the dynamic range of interest and that the sensor exhibit stable gain.

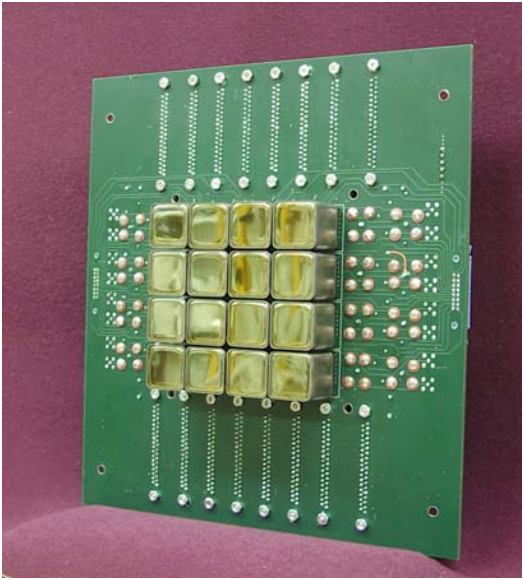

Figure 1: The TrICE camera board with 16 R8900 MAPMTs.

We considered several different MAPMTs, the Hamamatsu H8500 and the Hamamatsu R8900, and also considered the Burle micro-channel plate (MCP). The Burle MCP was rejected because the maximum average anode current rating was only 3uA, well below the large DC currents from the NSB and because of the relatively large dead area at the edges. We also rejected the Hamamatsu H8500 64-channel MAPMT although this tube is comparable to the R8900 for many of the parameters we studied. The H8500 failed to meet our single PE requirement. The Hamamatsu R8900 satisfied all of our critera. We next discuss each of these considerations separately.

**Single Photo-electron Requirement**

The bulk of our sensor chacterization work was performed in a test stand developed for studying photosensors. The test stand included an LED, a reference PMT, an x-y stager with an automated fiber positioner, and an exponentially-graded neutral-density filter connected to a stepper motor and DAQ readout. The electronic read out consisted of a 16-bit ADC and an amplifier board to yield 1.144fC/ADC charge resolution. The test stand is shown in Figure 2.

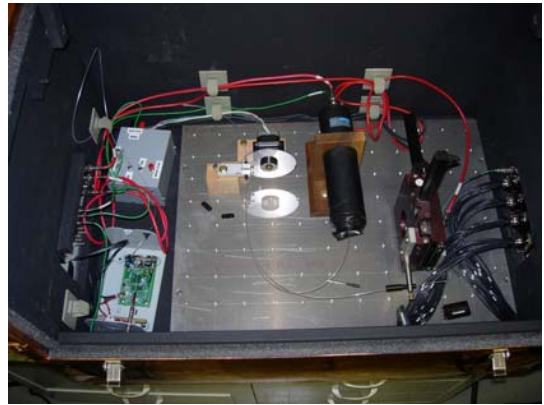

Figure 2: Photosensor characterization test stand.

The ability to observe a single PE peak in a photosensor is desired to track and understand the gain of the device. We were able to observe a single PE peak with both the MCP and the Hamamatsu R8900 MAPMT (Figure 3). This was not the case for the Hamamatsu 64-channel H8500.

**Low Crosstalk and Good Linear Response Requirement**

Low crosstalk between adjacent pixels and good linear response over the dynamaic range of 1 to 100 PE are other specifications for our photosensors. Both of these requirements are important to be able to accurately characterize shower distributions. Both the H8500 and R8900 MAPMTs showed low crosstalk in adjacent pixels, on the order of 2-3%, and good linear response over the dynamic range of 1 to several 100 photoelectrons. We did observe a deviation of ~5% from linearity at the highest levels, above approximately 200 to 500 PE.

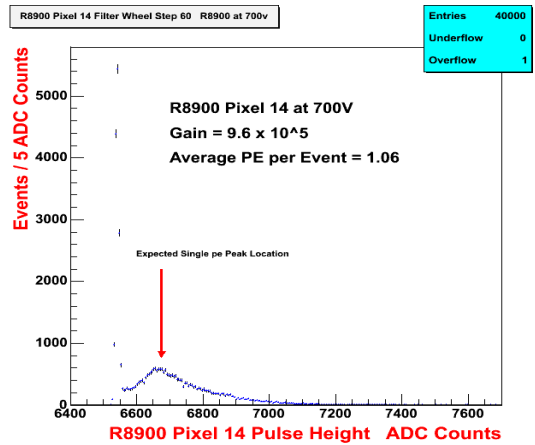

Figure 3: Single PE measurement of the R8900.

## Operation with Large DC Anode Currents with Stable Gain

One of the technological challenges of the TrICE camera was to determine if MAPMTs would have the sensitivity to measure signals of 3-100 photo-electrons in a shower in the presence of the high rates of photons from the NSB. For a 6 x 6 mm$^2$ pixel, the NSB rate is approximately 1 MHz for TrICE. These large rates required modifications of the base of the MAPMT to be able to operate the tubes at even moderate gains. The modification was to reduce the resistor components in the resistor chain of the base by 1/5. The currents measured at the TrICE site (very close to the Chicago metropolitan area) were still greater than 1 μA for a single MAPMT pixel with pitch of 6 mm operating at modest gain (~2 x 10$^5$). The large currents severely restrict operation of the phototubes under large NSB. The gain degrades noticeably at currents approaching 1 μA.

To test the effects of the NSB and base modifactions to compensate for it, a background was simulated using a DC light source in the test stand. This determined the optimal HV at which to operate our tubes. We used both an incandescent bulb to simulate the NSB and a blue LED to simulate a Cerenkov signal. A series of MAPMT gain measurements were made with the background light and compared to those with no background light. Figure 4 shows the results of this study as a function of the number of photoelectrons. Several curves are plotted in this figure; each color represents a unique DC illumination. These trends were similar for both the H8500 and R8900 Hamamatsu MAPMTs.

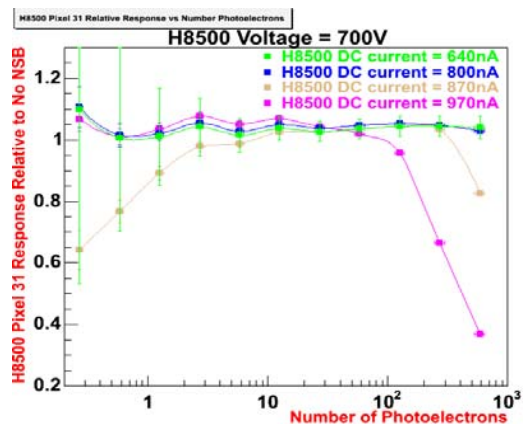

Figure 4: Relative gain response of an MAPMT with increasing intensity. This is shown for several simulated NSB levels similar to the conditions for TrICE.

The last specification for our camera was that the MAPMTs have stable gain to avoid the necessity for frequent tube calibrations. At the minimum, we required the gain be stable over the course of a night and this appears to have been the case. We were able to remove the gain-to-gain pixel variations with corrections derived from both data and LED calibrations. Figure 5 demonstrated the effect of our gain calibrations.

## Camera Calibration

The R8900 pixel-to-pixel gain varies by a factor of 2 to 3 within an MAPMT. The gain differences must be measured and corrected to properly interpret the images. Measurements of the relative gain are made by illuminating the camera uniformly with a difused LED.

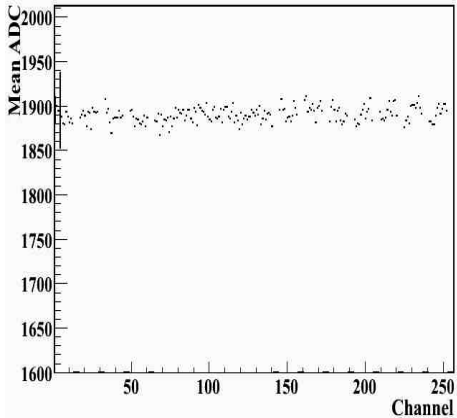

Figure 5: The mean gain-corrected charge is shown for each channel in ADC counts for a large number of events taken using uniform illumination from an LED.

Gain measurements were conducted on a nightly basis to monitor the response over time. Measurements made at different times throughout a single night show that intranight variation is dominated by the initial warm-up effects during the first 15 minutes after high voltage is applied to an MAPMT, and to a much lesser extent by temperature variations. Following the warm-up period, the gain response of the MAPMTs is stable.

## Camera Performance

The TrICE camera was completed earlier this year and commissioning is now mostly finished. Each PMT has undergone a suite of lab tests to characterize the gain for each pixel and to determine a baseline operating voltage. The voltage applied to each MAPMT was further adjusted once the camera was installed in TrICE to match the mean gain. In particular, the MAPMTs used in the trigger were selected to produce well-matched spectra for the dynode signals. Figure 6 displays an air shower image obtained using the TrICE MAPMT camera. Further results from TrICE are presented in [3].

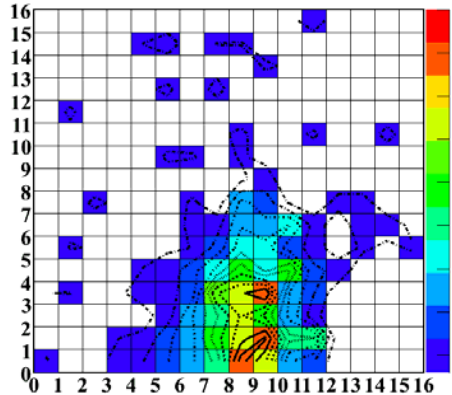

Figure 6: Cerenkov shower image in the TrICE camera. The scale is in relative charge. Pixels with small charge are suppressed in the figure.

## Conclusions:

In this paper we report on successfully developing a high resolution Cerenkov camera using MAPMTs. The camera has been used to image cosmic-ray air showers at higher angular resolution than current imaging Cerenkov telescopes. We present the first TrICE prototype images in this paper.